\documentclass[sigconf]{acmart}
\AtBeginDocument{%
  }

\copyrightyear{2026}
\acmYear{2026}
\setcopyright{cc}
\setcctype{by}
\acmConference[SIGIR '26]{Proceedings of the 49th International ACM SIGIR Conference on Research and Development in Information Retrieval}{July 20--24, 2026}{Melbourne, VIC, Australia}
\acmBooktitle{Proceedings of the 49th International ACM SIGIR Conference on Research and Development in Information Retrieval (SIGIR '26), July 20--24, 2026, Melbourne, VIC, Australia}
\acmDOI{10.1145/3805712.3808477}
\acmISBN{979-8-4007-2599-9/2026/07}

\usepackage{threeparttable}
\usepackage{booktabs} 
\usepackage{hyperref}
\usepackage{url}
\usepackage{graphicx}
\usepackage{amsmath}
\usepackage{amsfonts}
\usepackage{amsthm}
\usepackage{bbm}
\usepackage{hyperref}
\usepackage{tikz}
\usepackage{bm}
\usepackage{xcolor}
\usepackage{pgfplots}
\usepackage{multirow}
\usepackage{booktabs}
\usepackage{algorithmic}
\usepackage{algorithm}
\usepackage{array}
\usepackage{booktabs}
\usepackage{multirow}
\usepackage{amsmath}
\usepackage{subfig}
\usepackage{caption}
\usepackage{multirow}
\usepackage{algorithmic}
\usepackage{subcaption}
\usepackage{siunitx}
\begin{document}

\title{Towards Efficient and Generalizable Retrieval: Adaptive Semantic Quantization and Residual Knowledge Transfer}






\author{Huimu Wang}
\authornote{Equal contribution.}
\affiliation{%
  \institution{JD.com}
  \city{Beijing}
  \country{China}}
\email{wanghuimu1@jd.com}

\author{Xingzhi Yao}
\authornotemark[1]
\affiliation{%
  \institution{JD.com}
  \city{Beijing}
  \country{China}}
\email{yaoxingzhi1@jd.com}

\author{Yiming Qiu}
\authornote{Corresponding author}
\affiliation{%
  \institution{JD.com}
  \city{Beijing}
  \country{China}}
\email{qiuyiming3@jd.com}

\author{Qinghong Zhang}
\affiliation{%
  \institution{JD.com}
  \city{Beijing}
  \country{China}}
\email{zhangqinhong.6@jd.com}

\author{Haotian Wang}
\affiliation{%
  \institution{Harbin Institute of Technology}
  \city{Harbin}
  \country{China}}
\email{wanght1998@hit.edu.cn}

\author{Yufan Cui}
\affiliation{%
  \institution{Peking University}
  \city{Beijing}
  \country{China}}
\email{2401210230@stu.pku.edu.cn}

\author{Songlin Wang}
\affiliation{%
  \institution{JD.com}
  \city{Beijing}
  \country{China}}
\email{wangsonglin3@jd.com}

\author{Sulong Xu}
\affiliation{%
  \institution{JD.com}
  \city{Beijing}
  \country{China}}
\email{xusulong@jd.com}

\author{Mingming Li}
\authornotemark[2]
\affiliation{%
    \institution{Institute of Information Engineering, Chinese Academy of Sciences}
  \city{Beijing}
  \country{China}}
\email{limingming@iie.ac.cn}

\renewcommand{\shortauthors}{Wang, Yao, Qiu, et al.}

\renewcommand{\shortauthors}{Wang Huimu et al.}

\begin{abstract}
While semantic ID-based generative retrieval enables efficient end-to-end modeling in industrial applications, these methods face a persistent trade-off. On one hand, data-rich head items often suffer from ID collisions, which blur their distinct features and degrade downstream tasks. On the other hand, data-sparse tail items especially cold-start items are prone to semantic fragmentation during quantization; they are often mapped as isolated discrete points, which severely hinders their ability to generalize. To address this issue, we propose the Anchored Curriculum with Sequential Adaptive Quantization ($SA^2CRQ$) framework. The framework introduces Sequential Adaptive Residual Quantization (SARQ) to dynamically allocate code lengths based on item path entropy, assigning longer, discriminative IDs to head items and shorter, generalizable IDs to tail items. To mitigate data sparsity, the Anchored Curriculum Residual Quantization (ACRQ) component utilizes a frozen semantic manifold learned from head items to regularize and accelerate the representation learning of tail items. Experimental results from a large-scale industrial search system and multiple public datasets indicate that $SA^2CRQ$ yields consistent improvements over existing baselines, particularly in cold-start retrieval scenarios.
\end{abstract}




\maketitle

\section{Introduction}
\label{sec:introduction}

Generative Retrieval (GR) encodes a corpus into model parameters using a Sequence-to-Sequence architecture, enabling an end-to-end retrieval process \citep{se-DSI-tang2023semantic,seal,NCI,tome2023-url,li2023adaptive,li2023learning,li2023multiview,li2023generative,GenRet-baidu,yuan2024generative,sun2024learning,li2024generativeretrievalpreferenceoptimization,qiu2021query} in search and recommendation systems \citep{tiger,hou2025rpg,EAGER,enhancinggenerativeretrieval,lelter,onerec,MMGRec,adapticeCF}. Existing GR paradigms generally fall into two categories: lexical-based methods \citep{seal}, which use textual tokens as identifiers, and semantic ID-based methods \citep{tiger,NCI,zheng2025enhancingembeddingrepresentationstability,deng2025onerec,chen2025onesearch}, which utilize structures like clustering centers or residual codes. Given the strict performance requirements and the trend toward end-to-end modeling in real-world systems, semantic ID-based methods have gained wider adoption in industrial scenarios due to their superior inference efficiency and deployment feasibility.


\begin{figure*}[ht]
    \centering
    \includegraphics[width=1\linewidth]{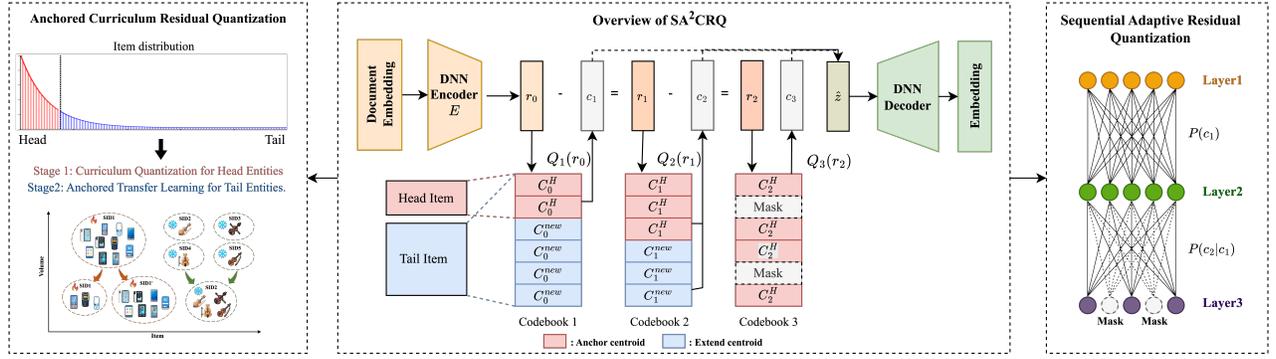}
    \caption{The framework of SA$^{2}$CRQ.}
    \label{fig:enter-label}
    \Description{Overview of the SA$^{2}$CRQ framework.}
\end{figure*}


However, deploying semantic ID methods in real-world applications reveals a fundamental structural trade-off driven by data imbalance. Empirical analysis of our large-scale e-commerce retrieval system, which contains billions of items, indicates that a mere 1\% of items account for approximately 79.34\% of historical interactions. This extremely long-tail distribution leaves millions of daily new arrivals and niche items highly data-sparse. Consequently, as illustrated in the lower-left part of Figure \ref{fig:enter-label}, the hyper-dense head items are forced to share semantic IDs, causing frequent collisions and blurred feature representations that degrade retrieval precision. Conversely, the massive volume of cold-start and tail items suffers from isolated, one-to-one mappings, limiting model generalization. 


To address the above issues, we propose the Anchored Curriculum with Sequential Adaptive Residual Quantization Framework ($SA^2CRQ$) (Figure \ref{fig:enter-label}). The framework systematically resolves the tension between representation discriminability and generalization through two synergistic mechanisms. First, we propose Sequential Adaptive Residual Quantization (SARQ) to determine the semantic ID length for each item dynamically. Guided by the information-theoretic principle of path entropy, SARQ assigns longer, high-fidelity codes to data-rich head items to increase discriminability, and shorter, generalizable codes to data-sparse tail items. Second, we introduce Anchored Curriculum Residual Quantization (ACRQ), a two-stage training strategy that facilitates structured knowledge transfer. ACRQ establishes a robust semantic manifold from head items, subsequently utilizing this manifold as a topological anchor to regularize and guide the representation learning for tail and cold-start items.

The main contributions of this work are:
\begin{itemize}
   \item We propose the \textbf{SA$^2$CRQ} framework, the first to unify adaptive length encoding and curriculum-based knowledge transfer to systematically address the long-tail challenge in generative retrieval.
  \item We introduce two synergistic components: \textbf{SARQ}, which uses path entropy to dynamically adjust code length, and \textbf{ACRQ}, a two-stage paradigm that transfers robust semantic knowledge from head to tail items.
  \item We conduct extensive experiments on public and large-scale industrial datasets, demonstrating that SA$^2$CRQ significantly outperforms state-of-the-art baselines, especially in improving tail-item coverage and cold-start performance.
\end{itemize}

\section{Methodology}
\label{sec:method}

To address the severe representation dichotomy between data-rich head items and data-sparse tail items in industrial systems, we introduce the Anchored Curriculum with Sequential Adaptive Quantization ($SA^2CRQ$) framework as shown in Figure \ref{fig:enter-label}. It builds upon the Residual Quantized Variational Autoencoder (RQ-VAE) \cite{tiger}, systematically resolving this tension through two core concepts: Sequential Adaptive Residual Quantization (SARQ) and Anchored Curriculum Residual Quantization (ACRQ), which are unified into a two-phase synergistic training paradigm.

\subsection{SARQ: Sequential Adaptive RQ}
Standard fixed-length codes force a suboptimal compromise: head items require longer codes to avoid ID collisions, while sparse tail items require shorter codes to prevent overfitting. SARQ dynamically determines the code depth based on the statistical uncertainty of the data. 

We quantify the information content of an encoding path using path entropy, decomposed via the chain rule in log-space:
$$\mathcal{I}(c_1:c_{l-1}) = -\log P(c_1) - \sum_{i=2}^{l-1} \log P(c_i | c_1, \dots, c_{i-1})$$
During training, SARQ frames quantization as a sequential decision process. At each layer $l \ge 2$, the path entropy is evaluated against an information budget $\mathcal{B}_l$: quantization proceeds to layer $l$ if $\mathcal{I}(c_1:c_{l-1}) \le \mathcal{B}_l$; otherwise, it terminates. Consequently, head items (residing in dense regions) consistently satisfy the budget and receive full-depth codes for maximal discriminability, while tail items exceed the budget and trigger early termination, constraining them to generic manifolds to enhance generalization.

\begin{figure*}[t!]
  \begin{minipage}{0.45\textwidth}
    \centering
    \includegraphics[width=\textwidth]{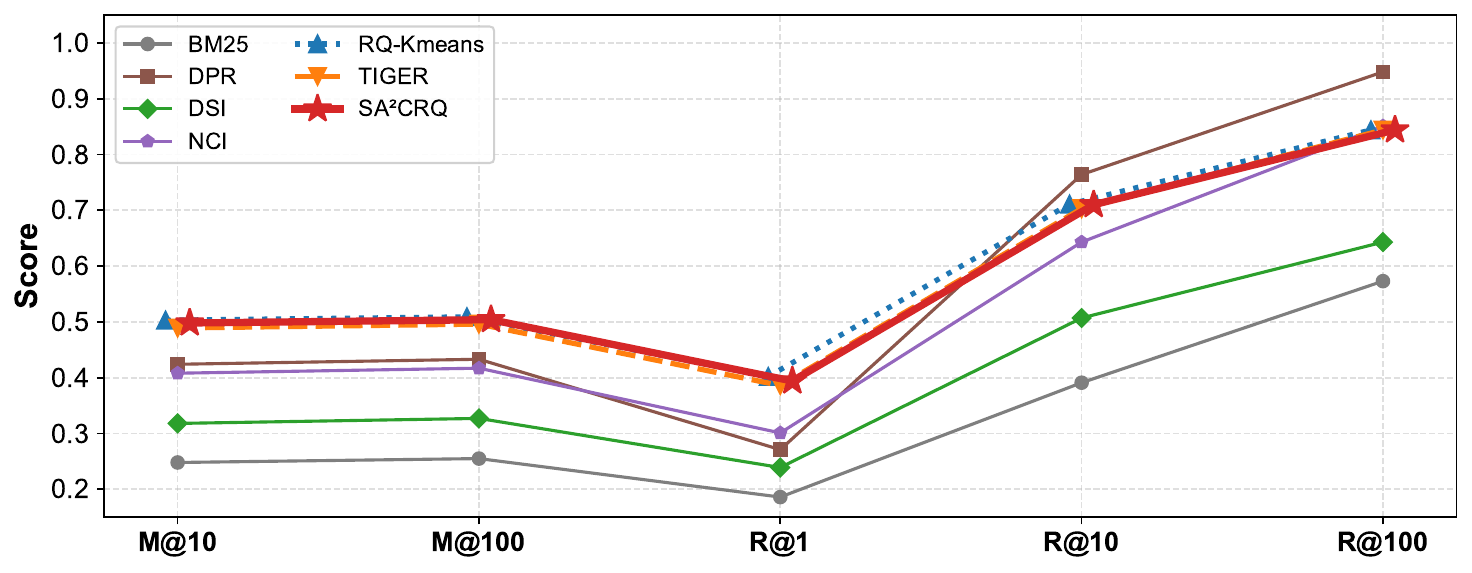} 
    \caption{Overall performance on Public Dataset.}
    \label{fig:with_aug}
  \end{minipage}
  \hfill
  \begin{minipage}{0.45\textwidth}
    \centering
    \includegraphics[width=\textwidth]{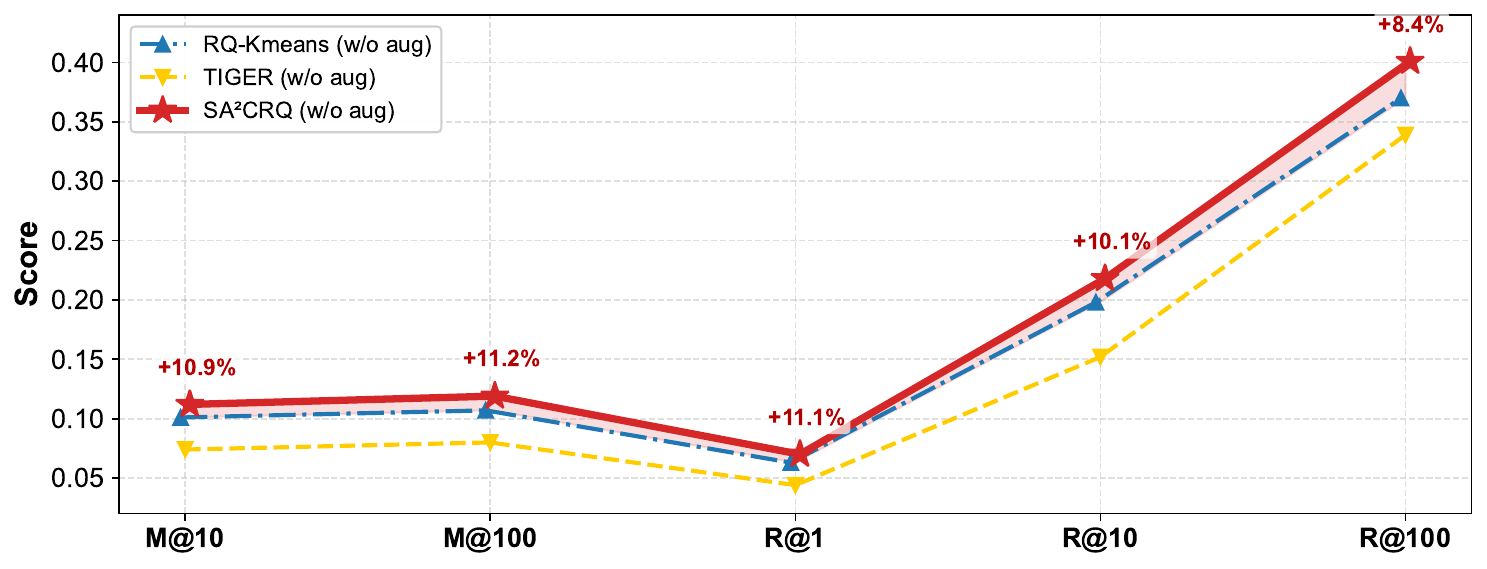}
    \caption{Performance under the sparse setting (\texttt{w/o aug}).}
    \label{fig:without_aug}
  \end{minipage}
  \label{fig:publicbench}
\end{figure*}

\subsection{ACRQ: Anchored Curriculum RQ}
ACRQ provides a principled mechanism for structured knowledge transfer to resolve the long-tail dichotomy, employing a two-stage paradigm.

\textbf{Curriculum for Head Items.} To capture macro-to-micro semantics, we approximate the Information Bottleneck principle by structurally constraining the codebook capacity. We enforce a monotonically increasing codebook size $M_l$ for each layer $l$ (e.g., $M_1 < M_2 < \dots < M_L = M$). A restrictive $M_1$ forces early layers to capture robust, low-entropy macro-semantics, while larger subsequent sizes encode fine-grained residuals. This phase yields a set of well-validated, hierarchically structured head codebooks $\{C_l^H\}$.

\textbf{Anchored Transfer for Tail Items.} Learning tail representations from scratch is prone to overfitting. ACRQ mitigates this by utilizing the established $\{C_l^H\}$ as a strong empirical Bayesian prior. We construct a hybrid codebook $C_l^T$ for each layer: $C_l^T = C_l^H \cup C_l^{\text{new}}$. The anchor part $C_l^H$ (size $M_l$) is strictly frozen, restricting the hypothesis space to established semantic regions, while the extension part $C_l^{\text{new}}$ (size $M-M_l$) remains trainable to capture novel, tail-specific patterns.

\begin{table}[t!]
\begin{threeparttable}
\centering
\small 
\caption{Performance of SA$^2$CRQ and its components on a large-scale industrial dataset.}
\label{indus_result}
\setlength{\tabcolsep}{2.5pt} 
\begin{tabular}{l ccccc}
\toprule
\textbf{Method} & \textbf{Hallu} $\downarrow$ & \textbf{R@500} $\uparrow$ & \textbf{R@1k} $\uparrow$ & \textbf{R@2k} $\uparrow$ & \textbf{Ret-Per$\dagger$} \\
\midrule
BM25 & - & 0.4712 & 0.5471 & 0.6271 & 2000 \\
\midrule
DPR & - & 0.5721 & 0.6753 & 0.7675 & 2000 \\
\midrule
RQ-Kmeans              & 0.2592 & 0.5938 & 0.6065 & 0.6105 & 401 \\
\quad (head)    & 0.2568 & 0.7260 & 0.7449 & 0.7518 & 459 \\
\quad (tail)    & 0.2624 & 0.4172 & 0.4209 & 0.4213 & 348 \\
\midrule
TIGER           & 0.2514 & 0.6212 & 0.6390 & 0.6448 & 509 \\
\quad (head)    & 0.2520 & 0.7440 & 0.7686 & 0.7772 & 572 \\
\quad (tail)    & 0.2497 & 0.4210 & 0.4283 & 0.4288 & 348 \\
\midrule
ACRQ            & 0.2225 & 0.6302 & 0.6583 & 0.6662 & 612 \\
\addlinespace[0.4em]
SARQ ($\tau={5e-7}$) & 0.1797 & 0.6137 & 0.6333 & 0.6388 & 555 \\ 
SARQ ($\tau={1e-6}$) & 0.1739 & 0.6183 & 0.6394 & 0.6453 & 579 \\
SARQ ($\tau={2e-6}$) & 0.1556 & 0.6451 & 0.6732 & 0.6798 & 661 \\
\midrule
SA$^2$CRQ ($\tau={5e-7}$) & 0.1734 & 0.6260 & 0.6665 & 0.6798 & 810 \\
SA$^2$CRQ ($\tau={1e-6}$) & 0.1519 & 0.6245 & 0.6713 & 0.6858 & 877 \\
SA$^2$CRQ ($\tau={2e-6}$) & \textbf{0.1808} & \textbf{0.6547} & \textbf{0.7061} & \textbf{0.7231} & \textbf{914} \\
\addlinespace[0.3em]
\quad (head)    & 0.1790 & 0.7085 & 0.7628 & 0.7820 & 940 \\
\quad (tail)    & 0.1890 & 0.4107 & 0.4484 & 0.4555 & 795 \\
\bottomrule
\end{tabular}
\begin{tablenotes}
    \footnotesize
    \item[$\dagger$] \textbf{Ret-Per} denotes the average number of retrieved items per query.
    \item[$\dagger$] \textbf{Hallu} denotes the proportion of generated semantic IDs that do not correspond to any real candidate item..
    \item[$\dagger$] \textbf{$\tau$} is the early-stopping threshold: if the item ratio under the first $N$ SID layers is no larger than $\tau$, quantization stops; otherwise, it proceeds to the next layer. We set $N=2$.
\end{tablenotes}
\end{threeparttable}
\end{table}

\subsection{SA$^2$CRQ: Unified Training Framework}
We unify the dynamic length allocation of SARQ (Section 2.1) and the structured knowledge transfer of ACRQ (Section 2.2) into the comprehensive SA$^2$CRQ framework. The standard RQ-VAE objective is adapted to seamlessly integrate these two components across a two-phase training paradigm. Let $k(x) \in [1, L]$ denote the dynamically determined code depth for an input $x$ derived from the SARQ mechanism.

\paragraph{\textbf{Stage 1: Head Training (ACRQ Curriculum + SARQ)}} The model is trained on the head dataset $D_H$ to construct the ACRQ hierarchical manifold. Simultaneously, SARQ dynamically truncates the quantization paths based on self-information. To align with this dynamic depth, the total loss $\mathcal{L}_H$ and its reconstruction component are explicitly conditioned on $k(x)$:
\begin{gather*}
    \mathcal{L}_H = \mathbb{E}_{x \sim D_H} \left[ \mathcal{L}_{\text{recon}}^{(k(x))} + \mathcal{L}_{\text{codebook}}^{(k(x))} + \beta \cdot \mathcal{L}_{\text{commit}}^{(k(x))} \right], \\
    \mathcal{L}_{\text{recon}}^{(k(x))} = \left\| \mathbf{z} - \sum_{l=1}^{k(x)} \mathbf{q}_l \right\|^2_2.
\end{gather*}
Constraining the summations up to $k(x)$ effectuates implicit regularization for items terminated early. This stage yields two critical outputs: the ACRQ structured head codebooks $\{C_l^H\}$ and the SARQ path probability tables $\{P_l^H\}$.

\paragraph{\textbf{Stage 2: Tail Training (ACRQ Anchored Transfer + SARQ Priors).}}
The model trains on the tail dataset $D_T$ using the ACRQ anchor-and-extend mechanism. The SARQ path termination $k(x)$ is determined by the head priors $\{P_l^H\}$ learned in Stage 1 to prevent representational drift. The total objective $\mathcal{L}_T$ is formulated similarly to $\mathcal{L}_H$. To update the learnable tail codebooks $\{C_l^{\text{new}}\}$ while keeping the anchor $\{C_l^H\}$ frozen, we apply gradient masking via an indicator function $\mathbb{I}(\cdot)$ in the codebook loss:
$$\mathcal{L}_{\text{codebook, T}}^{(k(x))} = \sum_{l=1}^{k(x)} \mathbb{I}(\mathbf{q}_l \in C_l^{\text{new}}) \cdot \|\text{sg}(\mathbf{r}_{l-1}) - \mathbf{q}_l\|^2_2$$
This masking restricts gradient updates to $C_l^{\text{new}}$ while maintaining the SARQ depth constraints.

\section{Experiments}
\label{sec:experiments}

\begin{figure*}[tp]
  \begin{minipage}{0.33\textwidth}
    \centering
    \includegraphics[width=\textwidth]{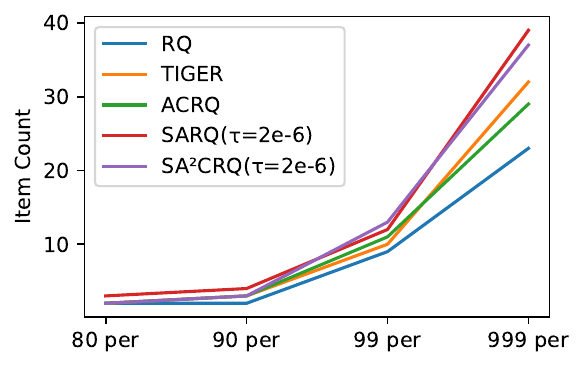} 
    \caption{The percent quantile of item count. RQ represents RQ-Kmeans.}
    \label{ITEMCONT-METHOD}
  \end{minipage}
  \hfill
  \begin{minipage}{0.31\textwidth}
    \centering
    \includegraphics[width=\textwidth]{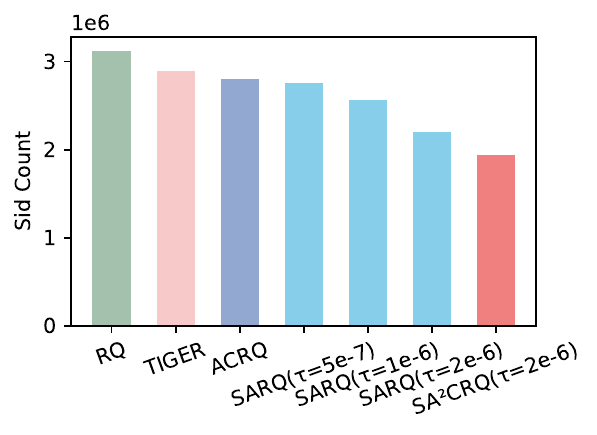}
    \caption{Total number of SIDs per method. RQ represents RQ-Kmeans.}
    \label{fig:sid_count}
  \end{minipage}
  \hfill
  \begin{minipage}{0.31\textwidth}
    \centering
    \includegraphics[width=\textwidth]{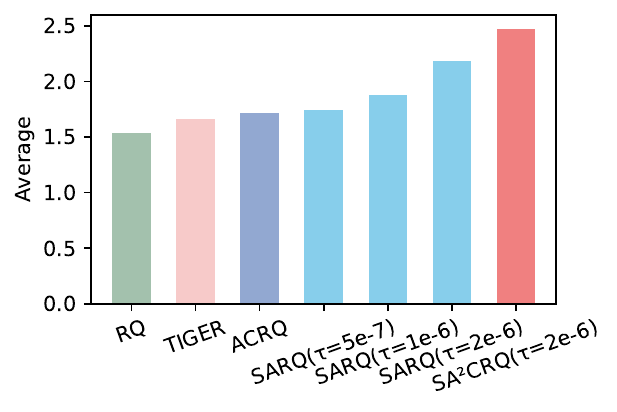}
    \caption{Average item count in SIDs. RQ represents RQ-Kmeans.}
    \label{fig:avg}
  \end{minipage}
  \label{fig:finaltotal}
\end{figure*}

\begin{figure*}[tp]
  \begin{minipage}{0.45\textwidth}
    \centering
    \includegraphics[width=\linewidth]{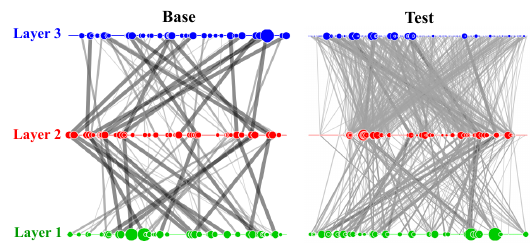}
    \caption{Comparison of distribution on head items.}
    \label{fig:distribution_one2many}

  \end{minipage}
  \hfill
  \begin{minipage}{0.45\textwidth}
    \centering
    \includegraphics[width=\linewidth]{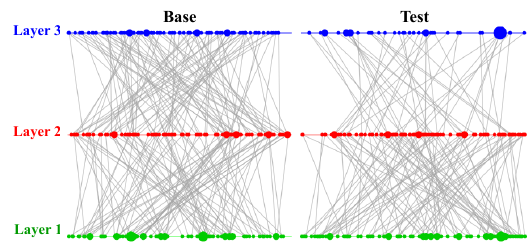}
    \caption{Comparison of distribution on tail items.}
    \label{fig:distribution_one2one}
  \end{minipage}
  \label{fig:finaltotal}
\end{figure*}

\subsection{Experimental Setup}
We evaluate SA$^2$CRQ on both an industrial e-commerce dataset from JD.COM and the public MS MARCO-derived MS300K benchmark. The JD.COM dataset contains \textbf{18M} queries and \textbf{5M} items, encompassing approximately \textbf{20M} training instances labeled by implicit click signals.This dataset preserves the native long-tail distribution without synthetic augmentation, and is further enriched with user behavioral sequences consisting of up to 20 historical clicks. MS300K is constructed from MS MARCO \cite{masmarco} following standard protocols \cite{novo,ultron,d2gen,tsgen,minder}. The dataset is split into training and test sets. During training, following the data augmentation strategy in NCI \cite{NCI}, we generate up to 10 pseudo queries for each document and perform deduplication to prevent queries in the test set from appearing in the training data.


To comprehensively evaluate system performance, we employ MRR@K (M@K) to measure ranking quality and Recall@K (R@K) to assess overall retrieval capacity \cite{tsgen,d2gen,DSI}. Furthermore, to explicitly gauge long-tail item coverage within the industrial setting, we introduce deeper recall metrics (R@500, R@1k, R@2k) and report the average number of retrieved items per query (Ret-Per). 

For a rigorous comparison, we select representative baselines across three dominant retrieval paradigms. Specifically, we compare against BM25 \cite{bm25} as the standard for sparse retrieval, and DPR \cite{dpr} for dense retrieval. Additionally, we evaluate against a suite of state-of-the-art generative retrieval methods, including DSI \cite{DSI}, NCI \cite{NCI}, TIGER \cite{tiger}, and the widely adopted semantic ID baseline RQ-Kmeans \cite{deng2025onerec,chen2025onesearch}.

\subsection{Implementation Details}
For experiments on the public benchmark, we use T5-base \citep{t5sentence} as the backbone model, which follows the Transformer encoder--decoder architecture. SA$^2$CRQ is trained for 200 epochs using the AdamW optimizer \cite{adamw}, with a learning rate of 5e-4 and a batch size of 1,024. Following prior baselines \cite{d2gen,tiger}, we adopt an RQ-VAE with 4 quantization layers and a codebook size of 256. During inference, the beam size is set to 100.

For the industrial e-commerce dataset, we use Qwen3-1.7B as the backbone model. The training process is conducted for 1 epoch with a learning rate of 2e-5. Considering the substantially higher item diversity in e-commerce scenarios compared with public benchmarks, we adopt a 3-layer quantization structure and increase the codebook size to 4,096. During inference, we perform beam search with a beam width of 128.

\subsection{Main Results on Industrial Search}
As shown in Table~\ref{indus_result}, our proposed $SA^2CRQ$ ($\tau=2e-6$) consistently outperforms all baselines on the industrial dataset. It achieves a Recall@2k of 0.7231, a 12.1\% relative improvement over the best generative baseline TIGER (0.6448), while reducing the hallucination rate from 0.2514 to 0.1808. These gains stem from the curriculum-guided anchored semantic space, which stabilizes decoding.


On tail items, $SA^2CRQ$ improves retrieval coverage (Ret-Per) from 509 to 914 and tail Recall@2k from 0.4288 to 0.4555. The dynamic length allocation in our approach assigns compact representations to sparse items, mitigating overfitting and enhancing zero-shot retrievability. Notably, head item performance is also preserved: by constructing a fine-grained semantic space via ACRQ, $SA^2CRQ$ avoids the identifier collision problem common in existing semantic ID methods, achieving a head Recall@2k of 0.7820 versus 0.7772 for TIGER. This demonstrates consistent retrieval quality across the entire data distribution.



\subsection{Generalization on Public Dataset}
Figure~\ref{fig:without_aug} evaluates the framework under authentic long-tail conditions without data augmentation. $SA^2CRQ$ consistently outperforms numerical-ID baselines, achieving R@100 of 0.401 (vs. TIGER's 0.339) and delivering relative gains of 8.4–11.2\% across all metrics. With augmentation (Figure~\ref{fig:with_aug}), it remains highly competitive (e.g., M@100=0.504), confirming robust representation across diverse data conditions.

\subsection{Ablation and Distributional Impact}


Ablation results (Table~\ref{indus_result}) indicate that isolating SARQ increases tail-item coverage (Ret-Per improves from 509 to 661). As shown in Figure~\ref{fig:distribution_one2one}, SARQ mitigates overfitting from fixed-length encoding by terminating tail-item paths early, yielding shorter adaptive codes. Furthermore, per-semantic-ID distributions (Figures~\ref{ITEMCONT-METHOD}, \ref{fig:sid_count}, and~\ref{fig:avg}) confirm that SARQ maps semantically related sparse items onto shared paths. This introduces controlled collisions among tail items, effectively trading fine-grained uniqueness for broader catalog retrievability.

Conversely, isolating ACRQ primarily improves overall precision, increasing R@2k from 0.6448 to 0.6662. Layer-distribution analysis (Figure~\ref{fig:distribution_one2many}) indicates that ACRQ mitigates the baseline's super-router problem for popular items. By establishing a hierarchical semantic manifold, ACRQ resolves one-to-many mapping bottlenecks and maintains head-item discriminability.

The full $SA^2CRQ$ model yields the best overall performance, indicating structural synergy: ACRQ provides a stable manifold for SARQ's path estimation, while SARQ guides sparse items toward these anchored paths. Hyperparameter analysis further confirms this control. As reported in Table~\ref{indus_result}, the budget $\tau$ regulates the generalization-precision trade-off, where stricter budgets encourage shorter codes to maximize tail coverage. Concurrently, restrictive capacities in early layers (e.g., the $[512,512,4096]$ split in Table~\ref{tab:head_tail_split}) further limit head-item collisions.

\begin{table}[t]
\centering
\caption{Sensitivity of head-tail codebook splits. \#SID: total semantic IDs; percentages: SID-item distribution quantiles; Codebook: head centroids per layer (total 4096).}
\label{tab:head_tail_split}
\setlength{\tabcolsep}{2.8pt}
\begin{tabular}{l c c c c c c c}
\toprule
\textbf{Codebook} & & \textbf{\#SID} & \multicolumn{4}{c}{\textbf{Quantiles}} & \textbf{Mean} \\
\cmidrule(lr){4-7}
Config. & & ($\times 10^6$) & 80\% & 90\% & 99\% & 99.9\% & \\
\midrule
{[$4096,4096,4096$]} && 2.897 & 2.0 & 3.0 & 10.0 & 32.0 & 1.6597 \\
{[$1024,2048,4096$]} && 2.917 & 2.0 & 3.0 & 10.0 & 29.0 & 1.6481 \\
{[$512,512,4096$]}  && 2.806 & 2.0 & 3.0 & 11.0 & 29.0 & 1.7130 \\
{[$2048,2048,4096$]} && 2.976 & 2.0 & 3.0 & 10.0 & 28.0 & 1.6153 \\
\bottomrule
\end{tabular}
\end{table}

\subsection{Online A/B Testing and System Efficiency}

To isolate and assess the real-world efficacy of our approach, $SA^2CRQ$ was evaluated on live traffic through a four-day strict ablation A/B test in JD.com's search engine, which serves hundreds of millions of users. The control group utilized the standard production RQ-Kmeans SID baseline without any of our proposed enhancements. 
Under this absolute ablation setting, the model achieved consistent positive lifts of \textbf{+0.13\%} in User Conversion Rate (UCVR) and \textbf{+0.42\%} in User Value.
At an industrial scale, these consistent lifts drive significant revenue, validating SARQ's adaptive tail-intent capture and ACRQ's head-item precision. Given its robust online performance in these isolated tests, $SA^2CRQ$ has now been \textbf{fully deployed} in the production system as part of a comprehensive version upgrade. 

In production, $SA^2CRQ$ demonstrates high efficiency: leveraging dynamic early termination to reduce decoding steps, our 1.7B-parameter model deployed on a single NVIDIA RTX 5090 sustains 30 QPS with 50ms TP99 latency and 30.4\% peak memory utilization, validating its feasibility for large-scale generative retrieval.


\section{Conclusion}
This paper introduces the $SA^2CRQ$ framework to resolve the fundamental trade-off between head-item discriminability and tail-item generalization in industrial generative retrieval. By synergistically integrating Sequential Adaptive Residual Quantization (SARQ) and Anchored Curriculum Residual Quantization (ACRQ), the framework dynamically allocates semantic ID lengths via path entropy while enabling structured knowledge transfer to data-sparse regions. Extensive offline evaluations and online A/B testing in a large-scale e-commerce system validate its practical effectiveness, notably doubling tail-item retrieval coverage and reducing hallucinations with manageable latency overhead. Future work will explore extending this adaptive encoding paradigm to multi-modal and cross-domain retrieval scenarios.

\bibliographystyle{ACM-Reference-Format}
\bibliography{sample-base}


\section*{Company Portrait}
JD.com, Inc., also known as Jingdong, is a Chinese e-commerce company headquartered in Beijing. It is one of the two massive B2C online retailers in China by transaction volume and revenue, a member of the Fortune Global 500. When classified as a tech company, it is the largest in China by revenue and 7th in the world in 2021.

\section*{Presenter profiles}
\noindent\textbf{Huimu Wang} is a researcher in the Smart Retail Department at JD.com Beijing. He received his doctor degree in Institute of Automation, Chinese Academy of Sciences. His research focuses on reinforcement learning and natural language processing.

\noindent\textbf{Xingzhi Yao} is an algorithm engineer in the Department of Search and Recommendation at JD.com Beijing. His research focuses on generative retrieval, with particular interests in semantic ID construction and post-training techniques for large language models.



\end{document}